\documentclass{moriond}

\def\be{\begin{equation}}
\def\ee{\end{equation}}
\def\bea{\begin{eqnarray}}
\def\eea{\end{eqnarray}}

\newcommand{\Photo}{}

\begin{document}
\vspace*{4cm}
\title{CHERENKOV TELESCOPE ARRAY STATUS REPORT}

\author{S. Mangano, on behalf of the CTA Consortium}

\address{CIEMAT - Centro de Investigaciones Energ\'eticas, Medioambientales y Tecnol\'ogicas\\ 
Av. Complutense, 40, 28040 Madrid, Spain}

\maketitle\abstracts{The Cherenkov Telescope Array (CTA) 
will be the next generation of ground based gamma-ray telescopes allowing us 
to study very high energy phenomena in the Universe.
CTA aims to gain about a factor of ten in sensitivity 
compared to current experiments, extending the accessible 
gamma-ray energy range from a few tens of GeV to some hundreds of TeV. 
This increased gamma-ray source sensitivity, as well as 
the expected enhanced energy and spatial resolution, 
will allow exciting new insights in some key science topics. 
Additionally, CTA will provide a full sky-coverage by featuring 
the array located in two sites in the Northern and Southern hemispheres.
This paper will describe the status of CTA and highlight
some of CTA's key science themes; namely 
the origin of relativistic cosmic particles, 
the study of cosmological effects on gamma-ray propagation and 
the search for annihilating dark matter particles.}

\section{Gamma-ray Astronomy with Cherenkov Telescope Array}
Gamma-ray astronomy is an exciting research field to 
study the very high energy phenomena in the Universe. 
The very energetic processes are opposed to thermal emission processes 
that originates from the random movements of particles with a given temperature. 
Many astrophysical objects exist that emit a significant fraction of 
non-thermal energy and many of them are the most explosive phenomena in the Universe, 
like Gamma Ray Burst (GRB)  and Active Galactic Nuclei (AGN).

The ground based observation of the most energetic phenomena in astrophysical objects 
through high energy gamma-rays is a young field.
The first discovery of an astrophysical object (Crab Nebula) 
at teraelectronvolt (TeV) energies~\cite{discoveryCrab} was achieved 
by the Whipple Observatory in Arizona in 1989. 
Since then, over 160 sources were detected 
demonstrating impressively the huge physics potential of this field and 
the maturity of the detection technique. 
However, it also became apparent that the performance of current instruments is not sufficient 
to exploit the rich physics potential. 
The answer of the high energy astronomy community to the need of a more sensitive and more flexible observatory 
is the Cherenkov Telescope Array (CTA)~\cite{CTAweb}.

CTA is designed to achieve an order of magnitude 
improvement in sensitivity in the energy range from $\sim 80$ GeV to $\sim 50$ TeV compared 
to currently operating instruments like VERITAS~\cite{VERITASweb}, MAGIC~\cite{MAGICweb} and H.E.S.S.~\cite{HESSweb}. 
These current telescope arrays host up to five individual telescopes, 
whereas CTA is expected to detect gamma-rays over a larger area 
with about 100 telescopes in the Southern hemisphere and about 20 telescopes in the Northern hemisphere.
CTA is also expected to have a much wider energy range coverage, 
a larger field of view, and a greatly improved energy and angular resolutions~\cite{CTABernloer}. 
These possible improvements will allow exciting new insights in key areas of astronomy, 
astrophysics and fundamental physics.

Due to the expected higher sensitivity the 
CTA observatory will detect more than one thousand sources 
allowing for population studies of source classes like AGN. 
Thanks to an improved angular resolution CTA will be able to perform 
very detailed morphological studies of extended sources.
As a result of a wider energy range coverage and improved energy resolution CTA 
will help to do spectral 
studies of unprecedented accuracy, giving stringent information about the acceleration mechanism.
Finally, owing to the large number of individual telescopes in the arrays, 
the observatory can operate with different pointing directions 
for different subsystems. For example some telescopes point to a distant galaxy and at the same time 
some other group of telescopes observe a Galactic source.

CTA will be the first open very high energy observatory, 
providing data products to the scientific community after some limited proprietary period.
CTA aims moreover to play a key role in 
multi-wavelength and multi-messenger astronomy with tight collaborations between
the different gravitational wave, neutrino, cosmic-ray, 
space and ground based electromagnetic telescope communities.

The CTA consortium is presently composed of 1350 members working in 210 institutes from 32 countries. 
These group of institutions are responsible for directing the science goals of the observatory 
and the array production.

\section{Imaging Atmospheric Cherenkov Technique}\label{sec:sites}
Gamma-rays interact with the Earth’s atmosphere and cannot be directly
detected on the terrestrial surface. Gamma-rays are therefore either directly observed from space using
detectors on satellites or indirectly from ground by detecting the electromagnetic
showers that are generated by gamma-rays interacting with the Earth atmosphere.

The typical astrophysical sources have a gamma-ray flux which is $dN/dE \approx E^{-2}$.
Therefore, the gamma-ray detection area has to be increased with gamma-ray energy to compensate 
for the fast power law decrease of the gamma-ray flux with increasing energy. 
Due to the low flux of gamma-rays above 100 GeV, detectors for such energies 
require a large detection area, ruling out space-based detectors 
with detection area of typical size of about 1~m$^2$. These {\it satellite telescopes} cannot
collect a statistically significant number of such high energy gamma-rays events 
in an reasonable amount of time. 
The gamma-ray {\it satellite telescopes} like {\it INTEGRAL}~\cite{Integralweb,Integralweb1}, \mbox{{\it COMPTEL}}~\cite{Comptelweb}, {\it Agile}~\cite{Agileweb} and {\it Fermi}~\cite{Fermiweb} are sensitive to energies 
of few tens of keV up to few hundreds of GeV. 
At higher energies two types of ground based gamma-ray instruments with 
different detection techniques exist. 
The first type of instruments observes the Cherenkov 
radiation that is produced by ultra-relativistic particles in an electromagnetic shower. 
Such instruments are called imaging atmospheric Cherenkov telescopes.  
The second type of instruments observes the particles
of the electromagnetic shower when they reach the ground level, like the HAWC~\cite{HAWCweb,HAWCweb1} detector. 

\begin{figure}
\centerline{\includegraphics[width=1.0\linewidth]{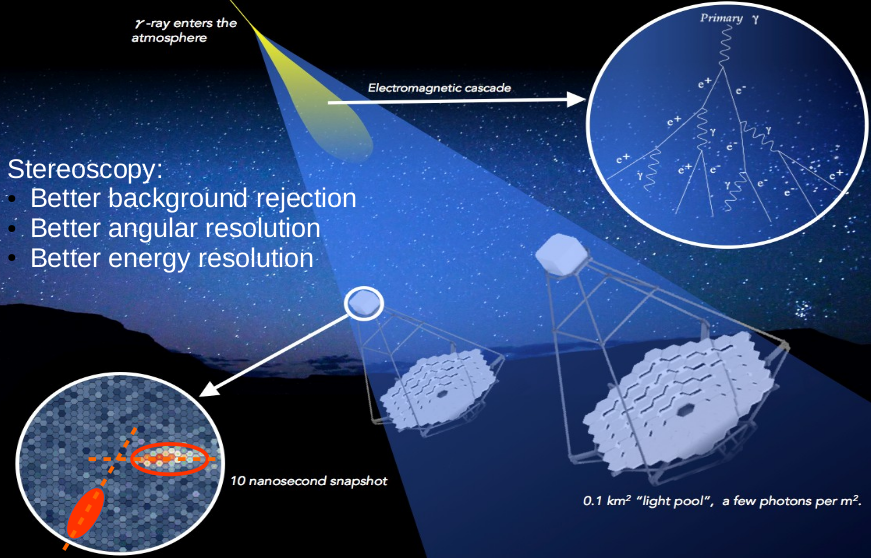}}
\caption[]{Schematic drawing of the imaging atmospheric Cherenkov technique. The first interaction of a very high energy gamma-ray with an atmospheric nucleus usually produces and electron-positron pair which in turn generates a cascade of particles, mostly electrons and positrons. The charged cascade particles emit Cherenkov light. A telescope located within the Cherenkov light pool can record an image of the primary gamma-ray. The simultaneous observation of a cascade with several telescopes (usually called spectroscopy) under different viewing angles increases substantially angular~\cite{Hofmann} and energy resolution.}
\label{fig:detecting}
\end{figure}

The aim of the imaging atmospheric Cherenkov telescopes is to take an image of the Cherenkov light produced 
in the Earth's atmosphere as presented schematically in figure~\ref{fig:detecting}. 
When a very high energy gamma-ray interacts with an atmospheric nucleus usually 
produces an electron-positron 
pair. The electron-positron pair in turn can 
interact with the atmosphere and generate a cascade of particles 
(usually called electromagnetic shower) which 
are mostly electrons and positrons of lower energy. The shower 
develop down in the atmosphere following certain general geometric properties. The shower typically 
measure several kilometers and have a width of some hundreds of meters. Some of the shower particles travel
at ultra-relativistic speed and emit Cherenkov light. This Cherenkov light propagates nearly 
unattenuated to the ground producing a faint pool of Cherenkov light of about 120 m in radius with a 
duration of a few nanoseconds. The effective detection area is at least the size of the pool of 
Cherenkov light on the ground, which is of the order of $10^5$~m$^2$. 
The optical mirrors of the telescopes reflect 
the collected Cherenkov light into the focal plane where a fast camera records the shower image.

The image in the camera represents the electromagnetic shower and is used to identify the primary 
cosmic gamma-ray. However, Cherenkov light is not 
only produced by gamma-rays but also by hadronic cosmic rays.
The shape, intensity and orientation of the image provides information about the primary particle type, 
energy and direction of the propagation. The elliptical shape of the gamma-ray image and 
the direction is the main feature to discriminate them 
against the overwhelming hadronic cosmic ray background which produce wider and more irregular images.

\section{Reconstruction Methods}\label{sec:reconstruction}
Reconstruction methods are used to separate the cosmic gamma-ray showers 
from the overwhelming hadronic cosmic ray showers.  
This separation is crucial for Cherenkov telescopes 
and is based on the different image pattern. 

The current methods used to separate signal 
and background in Cherenkov telescopes arrays are largely based on two approaches. 
The first approach uses the combination of an extended Hillas parameterization~\cite{Hillas} of the detected shower 
images with multivariate classification 
methods, like random forest~\cite{Randomforest}. The second approach fits the image to the result of a 
fast simulation under the hypothesis that the image is an electromagnetic shower~\cite{Mathieu,HESSplot}.
Although quite successful, both methods have some potential drawbacks. For the first one, 
the chosen parameters are not necessarily the best ones to deal with the signal and background 
separation problem, whereas for the second one the method is computationally very expensive.
 
Other approaches are currently under investigation, one of them is to study a 
deep network~\cite{deeplearning} 
to deal with the signal and background separation problem, and related problems, 
like energy and arrival direction reconstruction. 
On one hand deep network work by constructing an optimal and 
compact representation of the shower image that can be used to perform classifications. 
On the other hand deep networks work also as a 
reconstruction of the image under the assumption it belongs to the class 
of the image. 
Therefore deep networks seem ideal to merge the two approaches currently 
used with a computational cost during usage 
of the network close to the Hillas parameterization approach.

Currently machine learning techniques using TensorFlow~\cite{Tensorflow} 
are developed consisting of an image 
processing algorithm that can separate gamma-ray events from hadron events 
exploiting the full image recorded by the Cherenkov telescope on a pixel-wise information. 
A better classification of the images with 
respect existing methods could represent a major step forward in the field of observational gamma-ray 
astronomy.
Parallel to the the use of novel high level reconstruction strategies, the development of new hardware which is presented in section \ref{sec:concept},
can boost the scientific performance.

\section{Objective of Future CTA Project}\label{sec:sites}
The present Cherenkov telescope arrays have achieved very exciting results. 
These instruments are sensitive in an energy range range of $\sim 80$ GeV to $\sim 50$ TeV, have a typical
field of view of 3 degree to 5 degree, their angular resolution is about 0.05 degrees above 1 TeV and the 
energy resolution is about 10\% well above the threshold.
Nevertheless the future CTA projects should improve the performance of the 
existing instruments.

For example, no gamma-ray source has so far been detected by present telescopes 
in the energy domain above ∼100 TeV. Extending the observation in the ultra-high energy domain will
allow to understand the acceleration mechanism in galactic objects and help to 
 discriminate between hadronic and leptonic models~\cite{CTA1}.
On the other side of the energy spectra, lowering the energy threshold will be extremely
important for the observation of distant extragalactic sources because the gamma-ray 
flux at higher energies is attenuated by the interaction with the extragalactic background light.
In addition, a wide field of view combined with arc minutes
angular resolution will allow efficient surveys
and the study of extended sources. 

The main objective of the future CTA project is to allow very high energy gamma-ray astronomy 
to transit from source discovery to detailed source investigation. To reach this goal CTA will be an 
observatory aiming to:

\begin{itemize}
\item Extend the energy range from about 20 GeV to about 300 TeV, providing to detect high redshifts objects and extreme accelerators.
\item Increase the sensitivity by an order of magnitude at 1 TeV,  providing the possibility to detect many new sources.

\item Improve angular resolution down to few arc minutes, increasing rejection of hadronic background and improving the sensitivity.
\item Improve energy resolution, facilitating spectral studies to constrain acceleration mechanism.
\item Widen the telescope field of view, improving surveying capabilities and facilitate the study of extended sources.
\item Increase the detection area and therefore photon rate, providing access to observe fast transient phenomena.
\item Provide the access to the full sky.
\end{itemize} 

\section{CTA Concept with Mixed Size Telescopes}\label{sec:concept}

\begin{figure}
\centerline{\includegraphics[width=1.0\linewidth]{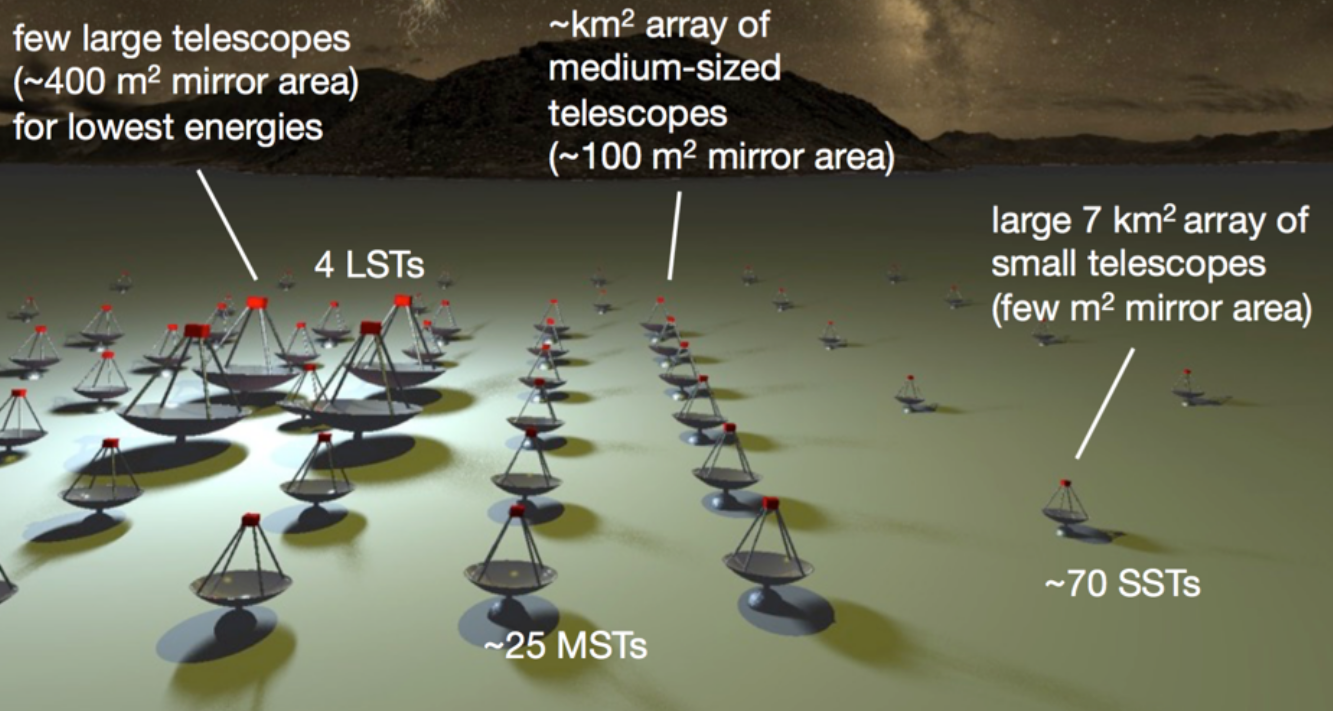}}
\caption[]{Artistic view (not to scale) of a possible layout of the CTA southern array consisting of three types of telescopes distributed over several square kilometers.}
\label{fig:schematic3types}
\end{figure} 

The CTA concept is artistically presented in figure  \ref{fig:schematic3types}
for a possible layout configuration.
To optimize cost rather than to put one type of a given size Cherenkov telescope on a regular distance,
the CTA concept consists in to use a mixed size telescope approach.
During the design phase the mixed size telescope approach has been optimized under 
the constraints given by cost limits 
and scientific requirements.

\begin{figure}
\centerline{\includegraphics[width=1.0\linewidth]{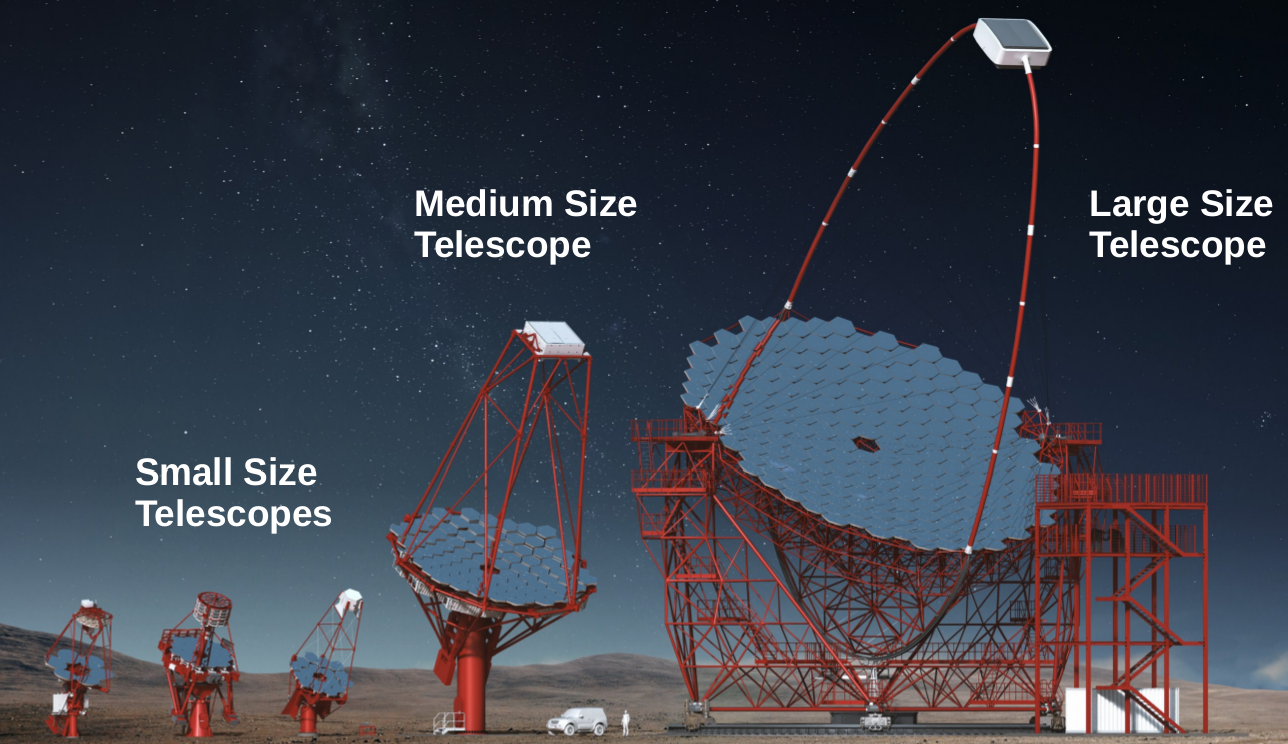}}
\caption[]{Schematic view of three different sized telescopes optimized for three different energy ranges. On the left three different small size telescopes are shown: the first two are based on the Schwarzschild-Couder reflectors and the next one is based on the Davis-Cotton optical design. The primary mirrors have a diameter of about \mbox{4 m}. In the middle a medium size telescope with a 12 m diameter Davis-Cotton mirror is depicted. Finally, on the right the design of the large size telescope with a parabolic dish of 23 m diameter is presented.}
\label{fig:3telescopes}
\end{figure}

CTA will be build with
three different size of telescopes to cover the full energy range in a cost effective way. 
The large sized
telescopes (LSTs) are optimized for energies from the lowest threshold to a few 100 GeV, 
the medium sized telescopes (MSTs) for the 
medium energy range from about 100 GeV to about 10 TeV, 
and the small sized telescopes (SSTs) for the highest energy range above a few TeV.
A schematic view of the sizes of the different telescope types can be appreciated in figure \ref{fig:3telescopes}.

In order to record a clean image of a gamma-ray shower,
a fast sampling of the signal is required for all telescopes.  The signal integration time has to be
as short as the Cherenkov light flash duration which is of the order of a few nanoseconds. 
All telescopes have in the focal-plane a camera, which use different numbers of very sensitive sensors 
for the detection of Cherenkov
light. The camera contains the readout electronics and it is 
connected to the data acquisition system by Ethernet optical fibers.

\subsection{Large Size Telescope}\label{sec:LST}

The LSTs are the largest CTA telescopes, and are 
optimized to detect the lowest energy gamma-rays. 
Since event rates are high and systematic effect of
the background limit the achievable sensitivity the detection area of these telescopes can be relatively small.
The low energies gamma-rays are efficiently detected by four closely placed
LSTs, with a mirror~\cite{mirror} diameter of about 23 m and a focal length of 28 m. Such large mirrors are needed 
to collect as many Cherenkov photons as possible from the low energy showers.
In order to discriminate the faint flashes 
produced by the lowest energy gamma-rays from the light of the night sky the 
LST camera~\cite{Delgado}  is the fastest and most sensitive one in CTA. The LST
will be supplied by a 1855 pixels camera with about 4 degree full filed of view.
The large mirror area together with up to 40\% quantum efficiency of 
the photomultipliers (PMTs) in the focal-plane camera and 
sophisticated trigger~\cite{trigger} and readout logic~\cite{readout} provides LSTs to detect gamma-rays 
down to 20 GeV.
This low energy range is 
extremely relevant because it allows the detection of 
the extragalactic gamma-ray 
sources like AGNs and of the sources with a limited acceleration 
power like pulsars. 

In addition, the LST should also allow the discovery of gamma-ray 
emission in GRBs. 
These telescopes require therefore a short repointing time to allow quick follow-ups of GRB alerts. 
The LSTs should be able to reposition to any direction in the sky in less than 20 
seconds after receiving an alert from a GRB monitor. In order to achieve a 
large size and light weight, the telescope structure is made of carbon fiber~\cite{carbonfiber} 
according to the approach followed by the MAGIC collaboration.
 
An artistic impression of the LST in comparison to the other telescopes is given in \mbox{figure \ref{fig:3telescopes}}.
The cost of each LST is about eight million euros and the first one will be also a 
prototype which will most probably be tested during 2018. 

\subsection{Medium Size Telescope}\label{sec:MST}
\begin{figure}
\centerline{\includegraphics[width=1.0\linewidth]{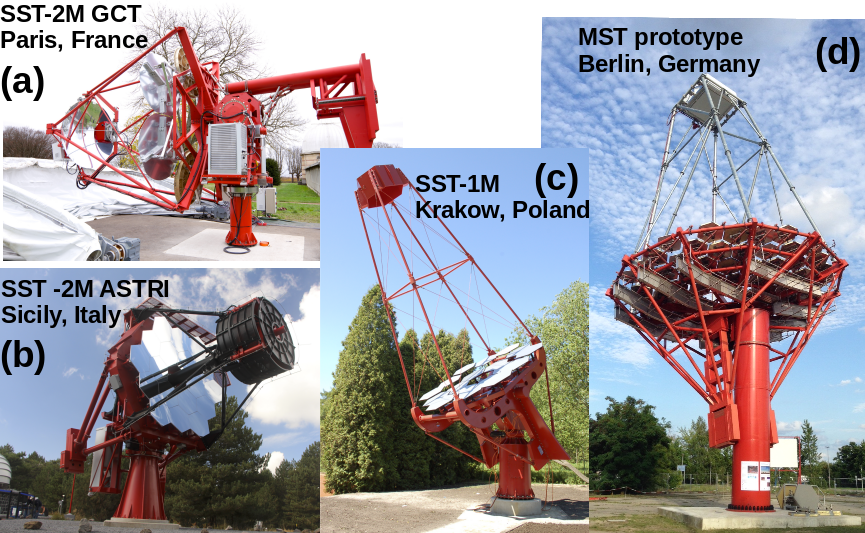}}
\caption[]{Photos of existing prototypes for the small size telescope (SST) and the medium size telescope (MST). Figure (a) shows the SST-2M GCT prototype in Paris and figure (b) the SST-2M ASTRI prototype in Sicily. Figure (c) presents the SST-1M prototype in Krakow and figure (d) the MST prototype in Berlin.}
\label{fig:prototypes}
\end{figure}

The energy range of about 100 GeV to about 10 TeV is covered by an array of MSTs with spacing 
of about 100 m. 
The shower detection and reconstruction in this energy range 
is based mostly on the 
experience of H.E.S.S. and VERITAS collaborations. The MST has a 
mirror diameter of about 12 m and a focal length of 16 m with a field of view of about 8 degrees~\cite{MST}. 
Improved sensitivity compared to current facilities will be obtained 
mainly due to increased area covered by the array
and by the larger number of telescopes reconstructing a shower.
The prototype MST shown in figure \ref{fig:prototypes}d is a modified Davies-Cotton telescope with a PMT-based camera.  
Two variants of signal readout are envisaged for the camera, 
the first one is the NectarCam~\cite{nectarcam} and the second one is the FlashCam~\cite{flashcam}. 
The NectarCam is a 1855 pixel camera sampling pulses with a
nominal frequency of 1 GHz using ASICs~\cite{asic}. 
The FlashCam is a 1764 pixel camera sampling signals 
with Flash-ADCs at a rate of 250 MHz and digital trigger.
In addition an alternative design is being developed, consisting 
on a dual-reflector Schwarzschild-Couder
configuration and silicon sensor camera comprising of 11328 pixels~\cite{Otte}. 
This design should have an improved performance, but is more technically challenging than the existing 
prototype MST.

\subsection{Small Size Telescope}\label{sec:SST}

The high energy range with showers with energy above 10 TeV are detected by an array of SSTs. 
The main limitation for the detection of high energy showers is the number of gamma-ray 
showers per time and area. 
Consequently, the array needs to cover an area of several square kilometers
to achieve a large statistic of high gamma-ray showers. 
At highest energies the light yield of a shower is large, so that showers can be detected 
even by small mirrors. Three different projects for the SSTs 
have been developed: two dual reflector solutions on the Schwarzschild-Couder optical system 
(SST-2M ASTRI~\cite{Ciro,Pareschi} and SST-2M GCT~\cite{Dournaux,Dournaux1}) and 
one based on single reflector on the Davis-Cotton optical design (SST-1M~\cite{Niemiec}). These three prototypes are 
presented in figure \ref{fig:prototypes}a, figure \ref{fig:prototypes}b and figure \ref{fig:prototypes}c.
The primary mirrors have a diameter of about 4 m and the field of view is about 9 degrees. 
The telescope ASTRI and SST-1M will be equipped with a silicon sensor camera~\cite{Heller}, whereas  
the GCT camera~\cite{Brown} can use either PMTs or silicon sensors.

\section{Full Sky-coverage and Possible Array Layouts}\label{sec:sites}
Access to the full sky is important as many of the objects to be studied by CTA are rare.
In order to have a full sky-coverage, the CTA observatory
will provide two arrays, with sites in both hemispheres.

The Northern hemisphere array will have an area of of radius 0.4 km and will be optimized 
in the energy ranges from $\sim $20 GeV 
to $\sim $20 TeV. This array is mainly dedicated to the observation of the northern extragalactic objects. Note that the absorption of high gamma-rays 
due to extragalactic background light can be large for extragalactic source 
so that no gamma-rays above few TeVs are expected.
The Southern hemisphere array with an area of radius 1.5 km will span the entire energy range, 
covering energies from $\sim $20 GeV to $\sim $300 TeV. A possible design 
for the southern site is presented in figure \ref{fig:schematic3types}.
The southern array will give the opportunity to observe the 
galactic sources, in particular the Galactic Center, and 
the southern extragalactic objects.



\begin{figure}
\begin{minipage}{0.5\linewidth}
\centerline{\includegraphics[width=1.\linewidth]{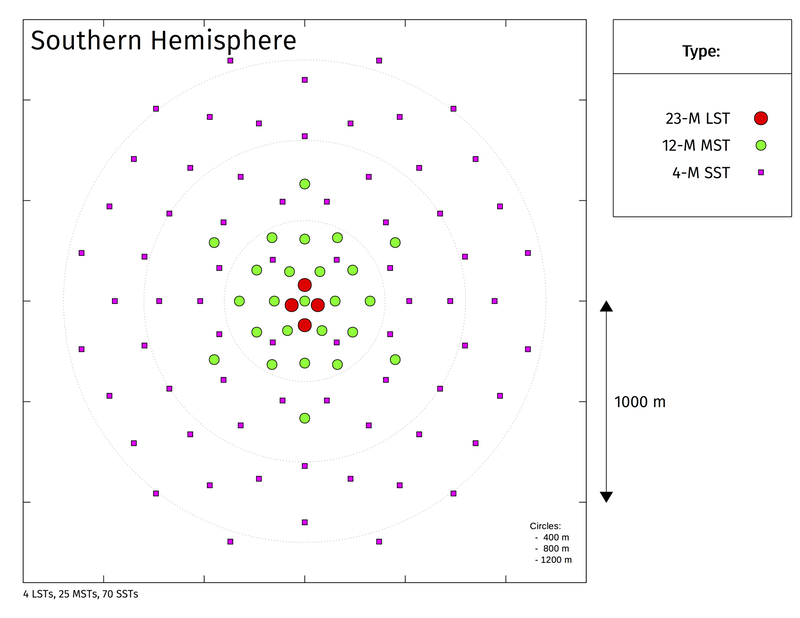}}
\end{minipage}
\hfill
\begin{minipage}{0.5\linewidth}
\centerline{\includegraphics[width=1.\linewidth]{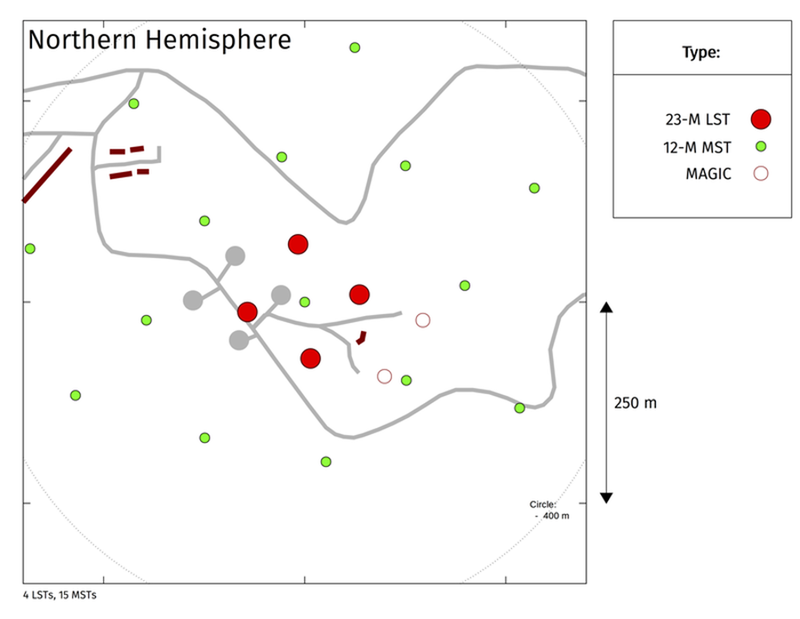}}
\end{minipage}
\caption[]{Left: Possible layout of telescope array in the Southern hemisphere. The southern array contains around 100 telescopes and will cover the full energy range from $\sim $20 GeV to $\sim $300 TeV. A southern array gives the opportunity to study the galactic sources and the southern extragalactic objects. Right: Possible layout of telescope array in the Northern hemisphere.  This array will be mainly dedicated to northern extragalactic objects and covers a smaller detection area than the southern array.}
\label{fig:posiblelayout}
\end{figure}


The plan for the southern site is to host  
the three sizes of telescopes to cover
the full energy rang, whereas the northern site will 
most probably be equipped with only LSTs and MSTs.
The Northern observatory will be made up of a 
19-dish array (4 LSTs and 15 MSTs) and 
a 99-dish array (4 LSTs, 25 MSTs and 70 SSTs) distributed over several square kilometers will be placed in the Southern observatory. 
The possible layout of the two CTA arrays are shown in 
figure \ref{fig:posiblelayout}. 
The expected lifetime of the observatory is about 30 years.

\section{Expected Performance of CTA}

\begin{figure}
\begin{minipage}{0.45\linewidth}
\centerline{\includegraphics[width=1.05\linewidth]{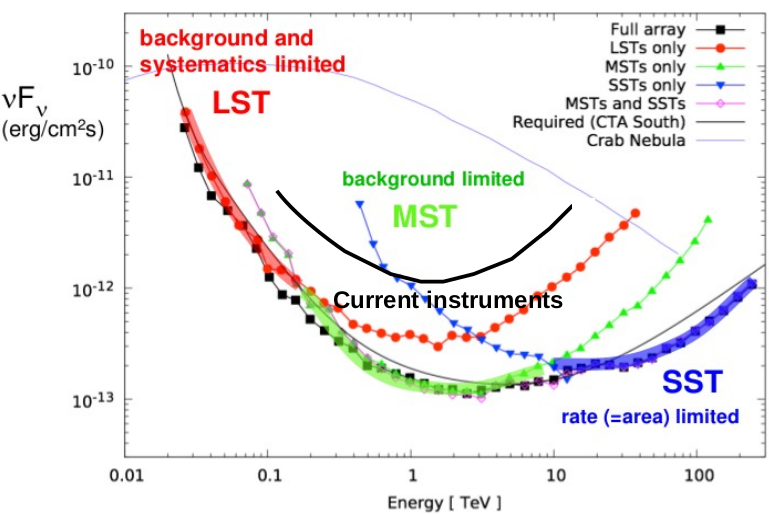}}
\end{minipage}
\hfill
\begin{minipage}{0.55\linewidth}
\centerline{\includegraphics[width=0.99\linewidth]{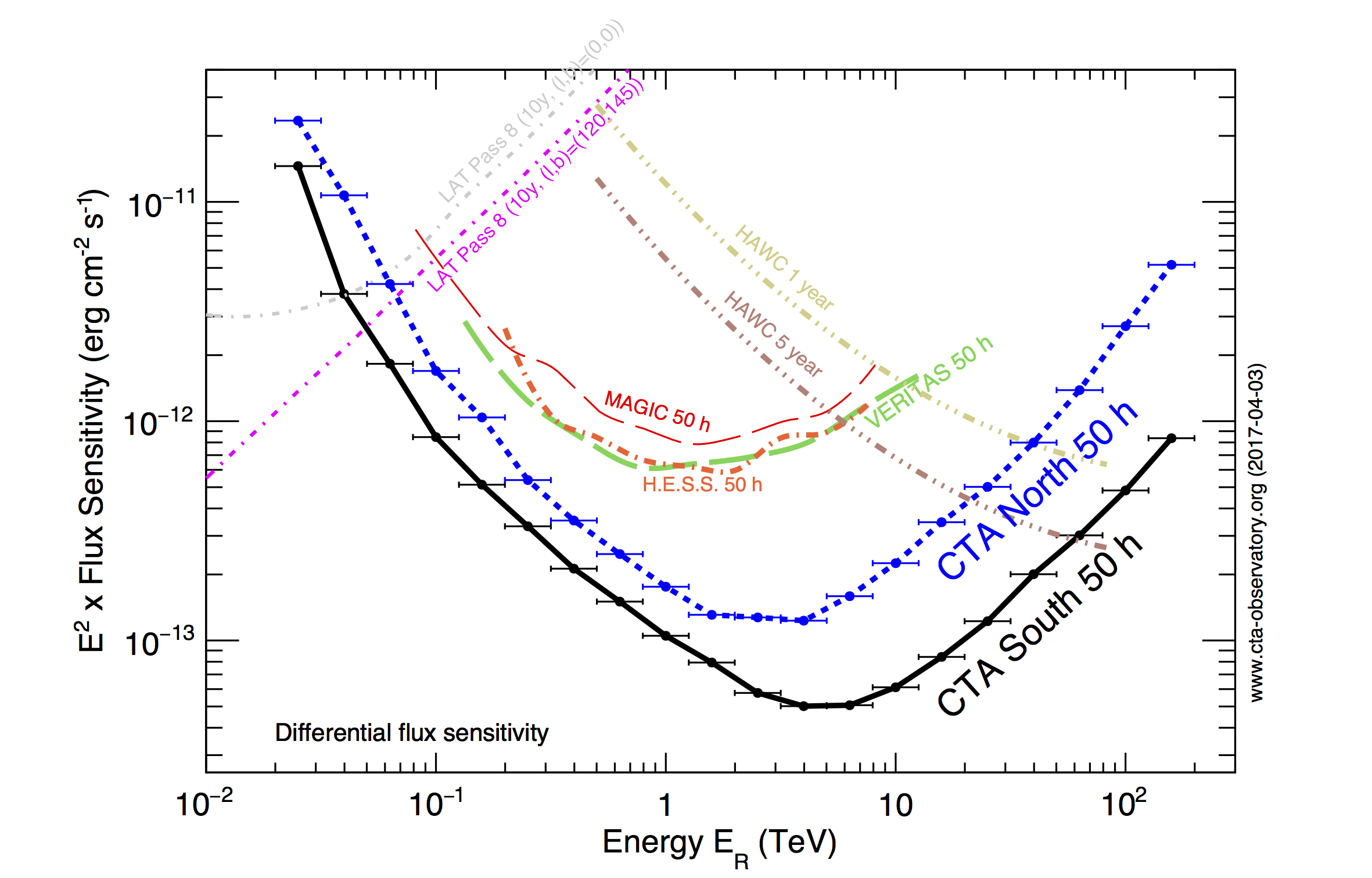}}
\end{minipage}
\caption[]{Left: Point source sensitivity for the full array layout (black line, filled squares). The full array sensitivity is compared to that of the different telescope types: 4 LSTs (red line, filled circles), 25 MSTs (green line, filled triangles), and 75 SSTs (blue line, upside down triangles). For comparison the current instruments sensitivity for the same observation time is shown (black line). Note that evolution of the sensitivity is expected as analysis algorithms improve and the final layout is fixed. Right: Point source differential flux sensitivity for CTA southern array (black solid line) and CTA northern array (blue dotted line). For comparison H.E.S.S.~\cite{HESSplot}, VERITAS~\cite{VERITASplot} and MAGIC~\cite{MAGIC} sensitivities for the same observation time are shown. Also presented is the HAWC sensitivity~\cite{HAWC} for an observation time of one year and five years, together with 10 years {\it Fermi}-LAT~\cite{Fermiplot} sensitivity with two different levels of diffuse gamma-ray background radiation.}
\label{fig:sensitivity}
\end{figure}

The expected performance of CTA has been derived through extensive Monte Carlo simulations 
taking into account many different possible array layouts~\cite{CTABernloer}. 
The expected derived results 
are produced by the 
so-called prod 2 version of the Monte Carlo simulations and up to date results can be found in the following website~\cite{CTAperformance}.

The contribution of the full array together 
with the different types of telescopes to the sensitivity for point sources is presented in the 
figure \ref{fig:sensitivity} left.  
The improvements in sensitivity and the wider energy range with respect to current instruments are clearly visible.
The sensitivity is computed requiring $5~\sigma$ detection in 50 hours 
of observation time with at least 10 gamma-rays, reaching a
sensitivity of a few milli Crab (1 Crab Unit = $2.79 \cdot 10^{-7}$m$^{-2}$s$^{-1}$TeV$^{-1} \times ($E/TeV$)^{-2.57}$).
The cross-over in sensitivity between LSTs and MSTs is seen at about 250 GeV and that between MSTs and SSTs is at about 4 TeV.
At these cross-over points, the combined sensitivity is almost a factor 
of two better than that of the individual components.
For low energies, the rejection of hadronic showers
is poor and the background dominates the signal.
For high energies, the backgrounds are at a
very low level and the point source sensitivity is only signal limited, 
due to the low rate of signal events.

The expected point source differential sensitivity for both southern and northern arrays for an
observation time of 50 hours is compared in figure  \ref{fig:sensitivity} right.
Also shown are
the sensitivities of MAGIC, VERITAS and H.E.S.S. for the same observation time. HAWC sensitivities are shown for an observation time of one year and five years, together with ten years {\it Fermi}-LAT sensitivity. The presented sensitivities give only an indicative comparison of the sensitivity of the different instruments, as the method of calculation and the criteria applied are different.  
The sensitivity of CTA southern array will be better than that of the northern side due to the higher number of telescopes and larger detection area. At energies above few TeVs the sensitivity of the southern array is increased due to 
inclusion of the SSTs. Nevertheless both arrays will outperform all existing
detectors over the full energy range from about 50 GeV to about 100 TeV.

At the lowest energies and for steady sources, CTA will be worse than 
the {\it Fermi}-LAT sensitivity~\cite{Fermicta}, due to that {\it Fermi}-LAT has a much higher background rejection. However, for short-time phenomena, such as gamma-ray bursts or AGN flare, CTA will be several orders
of magnitude more sensitive than {\it Fermi}-LAT even at lower energies, due to the 
larger effective detection area of CTA compared to {\it Fermi}-LAT. It is expected that 
CTA will be an ideal tool for transient phenomena,
probing sub-minute-timescale variability.

\begin{figure}
\begin{minipage}{0.5\linewidth}
\centerline{\includegraphics[width=1.\linewidth]{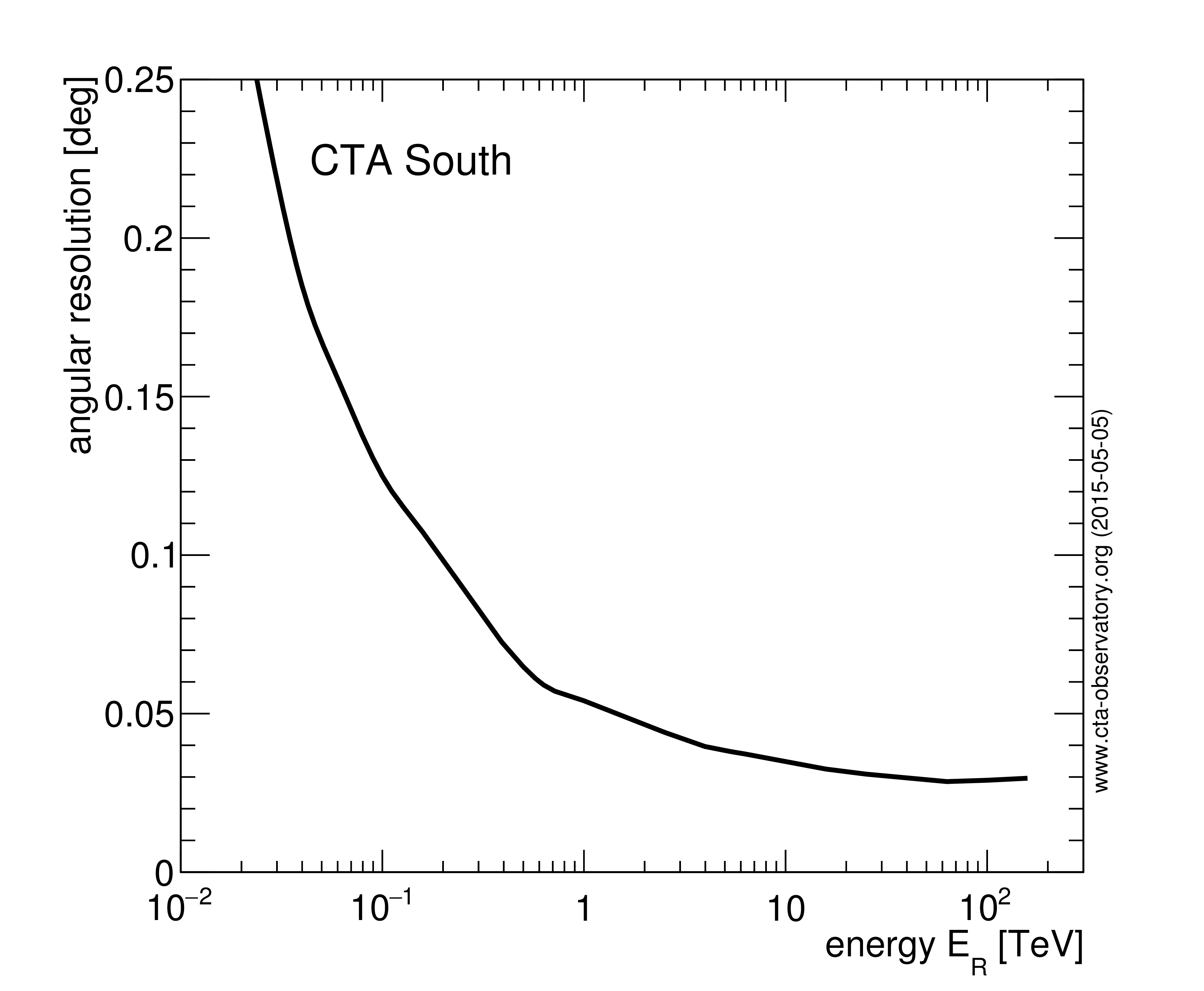}}
\end{minipage}
\hfill
\begin{minipage}{0.5\linewidth}
\centerline{\includegraphics[width=1.\linewidth]{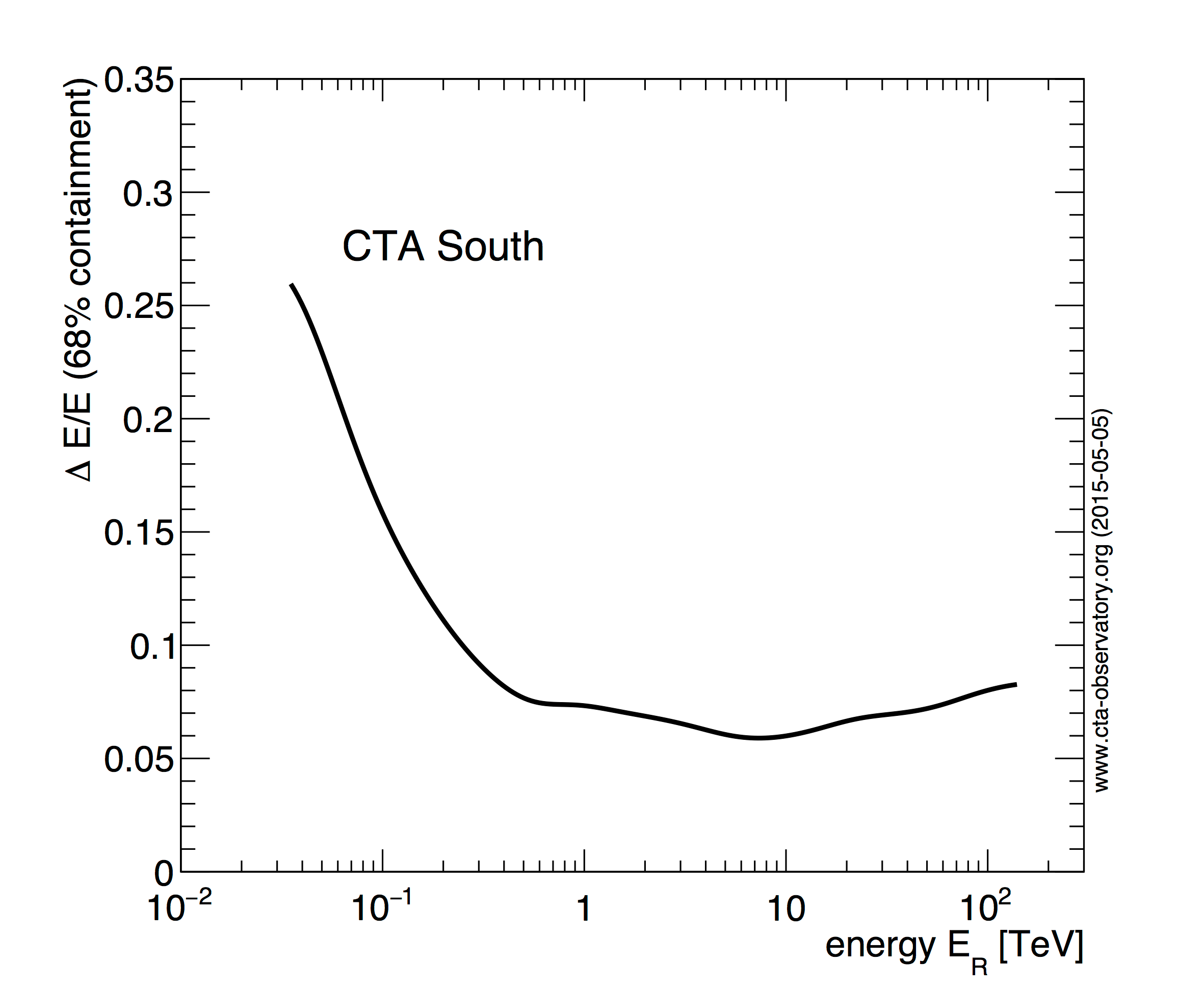}}
\end{minipage}
\caption[]{Left: Angular resolution (68\% containment radius of the gamma-ray point spread function) as a function of the reconstructed energy for the CTA southern array.  Right: Energy resolution $\Delta(E)/E$ of the CTA southern array as a function of reconstructed energy. Note that both presented results are not optimized to provide best resolutions, but rather best point source sensitivity.}
\label{fig:performance}
\end{figure}

In present arrays with at most five telescopes spaced by about 100 m, most 
of the detected showers have impact points outside the footprint of the array. The consequence is that the showers are detected usually by only one or two telescopes giving a modest angular resolution. This resolution can be improved
by selecting showers with larger number of telescopes. By increasing the number of telescopes over a larger areas than the size of the Chrenkov light pool the showers 
are contained inside the telescope array.
By  selecting  gamma-ray showers detected simultaneously by many of the telescopes, 
CTA can reach angular resolutions of better than 0.05 degrees
for energies above $\sim$1 TeV as seen in figure \ref{fig:performance} left. Note however that for this figure the result is not optimized to provide best angular resolution, but rather best point source sensitivity. 
Therefore higher resolution is possible at the expense 
of some detection area.  

 Figure \ref{fig:performance} right shows the energy resolution defined as the half width of the $\pm34~\%$ interval around the most probable reconstructed energy, divided by the most probable reconstructed energy. 
The presented expected performance are based on the combination of Hillas parameterization and multivariate classification methods.
Some improvement 
is expected with the use of more sophisticated techniques fully 
exploiting pixel-wise information as mentioned in section \ref{sec:reconstruction}.

\section{Science Themes}
CTA will explore questions in physics of fundamental importance, 
with considerable potential for major discoveries in astrophysics.
Some selected science themes of the CTA observatory 
are: 

\begin{enumerate}
\item Understand the origin of relativistic cosmic particles. 
\item Understand the role of the relativistic particles play in the evolution of star forming systems and galaxies.
\item Probe extreme environments such as neutron stars, black holes as well as cosmic voids.
\item Explore frontiers in physics such as the nature of dark matter.
\end{enumerate}

\begin{figure}
\begin{minipage}{0.55\linewidth}
\centerline{\includegraphics[width=1.\linewidth]{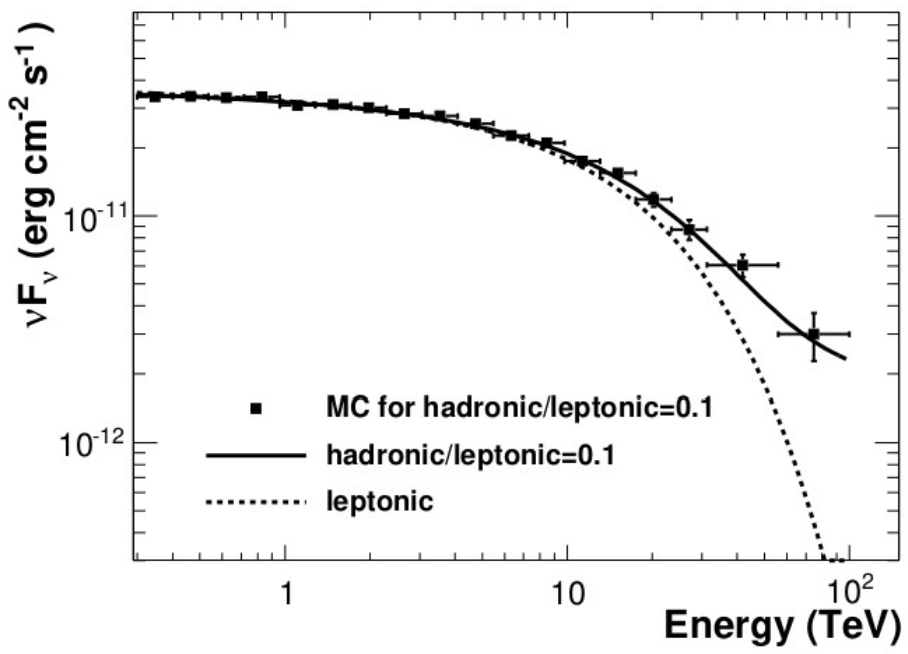}}
\end{minipage}
\hfill
\begin{minipage}{0.45\linewidth}
\centerline{\includegraphics[width=1.\linewidth]{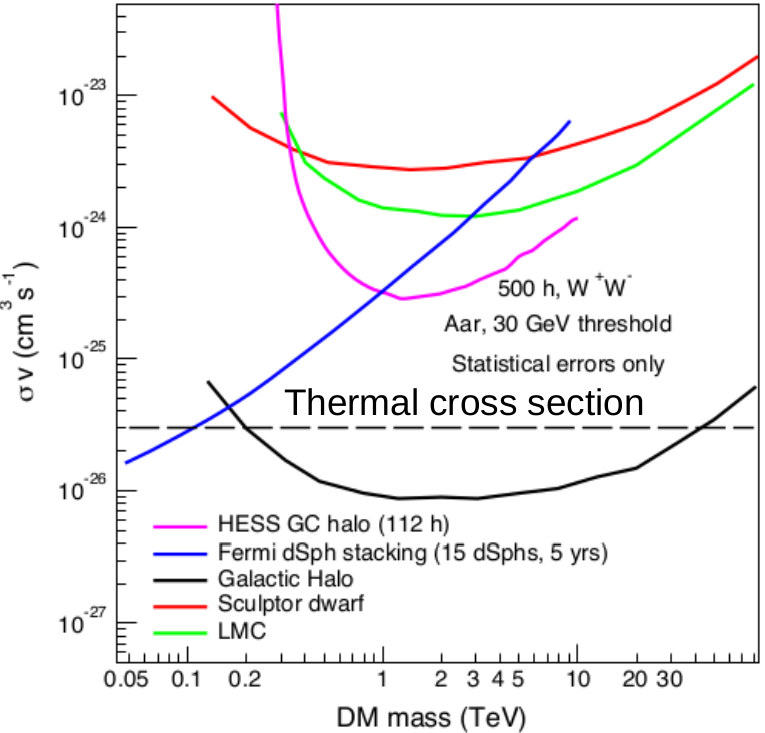}}
\end{minipage}
\caption[]{Left: Comparison of the spectral energy distribution from different gamma-ray emission. The black squares are the total of the fluxes for a leptonic and hadronic emission. The solid line shows the input spectra of gamma-ray simulation. The dotted line is for the case when the emission is purely leptonic. Right: The annihilation cross section $ \langle \sigma v \rangle $ as a function of dark matter mass. The measured cross section sensitivities of H.E.S.S.~\cite{hessdm} (pink) and {\it Fermi}-LAT (blue) are compared to the CTA expected sensitivities curves for Galactic Halo (black), dwarf galaxy Sculptor (red) and Large Magellanic Cloud (green) using 500 hours of observation. The predicted curves have been calculated using the Navarro-Frenk-White dark matter profile and W$^+$ W$^-$ annihilation modes. One can see that, CTA will reach the expected gamma-ray emission from annihilation of dark matter particles with masses from a few 100 GeV to few TeV, provided that their annihilation cross section corresponds to the thermal relic density value.}
\label{fig:physics}
\end{figure}

The science themes of the CTA array are described in detail in the ``Science with CTA'' document~\cite{CTAscience}, which is about to be published. There is a plethora of CTA science and only some selected 
results have been shown during the presentation, like the study of galactic and extragalactic acceleration mechanism as well as the observation of the long and short time variable phenomena. Figure \ref{fig:physics} left shows the expected spectra for different emission scenarios simulated for a typical galactic source~\cite{SNRcta}. The curves show a clear 
discrepancy detected at high energies according to different gamma-ray emission. The ambiguity between
leptonic and hadronic processes can be nearly completed resolved at gamma-ray energies around 100 TeV.
This is due to the Klein-Nishina effect, where leptonic
emission is highly suppressed at energies above few tens of TeV. 

In addition, CTA is perfectly suited to study cosmological 
effects on gamma-ray propagation. 
Results of different studies to probe 
extragalactic background light, intergalactic magnetic field, axions and
Lorentz invariance violation have been presented~\cite{CTAscience}. 
The main point is that by studying several sources at different distances, 
one can better disentangle intrinsic properties from photon propagation properties. 

Finally, many theories predict that dark matter particles could annihilate each other and 
emit gamma-rays that CTA should detect and expected sensitivities from such 
indirect dark matter searches have been presented. 
The indirect search method looks for gamma-ray emission 
in astrophysical regions with a high dark matter density.
The balance between the strength of 
expected dark matter annihilation signal, its uncertainty,
and the strength of the astrophysical backgrounds 
drives the astrophysical region selection.
Figure \ref{fig:physics} right shows the annihilation cross section 
$ \langle \sigma v \rangle $ as a function of dark matter mass~\cite{Carr}.  
The capability to discover a thermal weakly interacting massive particle, 
with the annihilation cross section 
$ \langle \sigma v \rangle = 3\times 10^{-26}$~cm$^3$~s$^{-1}$  is within reach with a deep exposure 
of the Galactic Halo with 500 hours. 

\section{Conclusion}
The achievements of the very high energy phenomena in the
Universe over the last three decades, 
fully justify the further exploration of the sky at these energies.
The answer of the very high energy astronomy community is 
the Cherenkov Telescope Array, a next generation, more sensitive 
and more flexible facility.
CTA, with a sensitivity increase of a factor about ten 
with respect to current telescopes, 
will  reveal  thousands  of  very  high  energy  gamma-ray  sources.
One of the key features of CTA will be to extend the energy
range from a few tens of GeV to above \mbox{100 TeV}. The construction of the first
telescopes has already started and working prototypes for small and medium 
size telescopes exist. CTA will operate as open astronomical observatories 
with exciting science in both hemisphere. 


\section*{Acknowledgments}
This paper has gone through internal review by the CTA Consortium.
I gratefully acknowledge the support of the project (reference AYA2014-58350-JIN) funded by MINECO through the young scientist call (year 2014).
This work is partially supported by the Maria de Maeztu Units of Excellence Program (reference MDM-2015-0509).
CTA gratefully acknowledges support from the agencies and organizations listed under Funding Agencies at this
website: http://www.cta-observatory.org/.


\end{document}